\documentclass[sigconf]{acmart}

\usepackage{hyperref} 

\usepackage{macros} 
\usepackage{booktabs} 
\usepackage{graphicx}
\usepackage{amssymb}
\usepackage{amsmath}
\usepackage{xfrac} 
\usepackage{booktabs} 
\usepackage{comment}
\usepackage{lineno}
\usepackage{todonotes} 
\usepackage[ruled,linesnumbered]{algorithm2e} 
\usepackage{multicol,multirow,dcolumn,longtable,footnote}
\usepackage{siunitx}
\usepackage{listings}
\usepackage{xspace}
\usepackage[flushleft,para]{threeparttable} 
\usepackage{ifthen}

\newcolumntype{P}[1]{D{.}{.}{#1}} 
\newcolumntype{V}[1]{>$ {\vcenter\bgroup\hbox\bgroup}#1<{\egroup\egroup} $} 

\newif\ifappendix  
\appendixtrue

\begin{document}

\copyrightyear{2019}
\acmYear{2019}
\setcopyright{author}  
\acmConference[HSCC '19]{22nd ACM International Conference on Hybrid Systems: Computation and Control}{April 16--18, 2019}{Montreal, QC, Canada}
\acmBooktitle{22nd ACM International Conference on Hybrid Systems: Computation and Control (HSCC '19), April 16--18, 2019, Montreal, QC, Canada}
\acmPrice{15.00}
\acmDOI{10.1145/3302504.3311804}
\acmISBN{978-1-4503-6282-5/19/04}

\title{JuliaReach: a Toolbox for Set-Based Reachability}

\author{Sergiy Bogomolov}
\affiliation{%
\institution{Australian National University}
\city{Canberra}
\country{Australia}
}

\author{Marcelo Forets}
\affiliation{%
  \institution{CURE - UdelaR}
  \city{Maldonado}
  \country{Uruguay}
}

\author{Goran Frehse}
\affiliation{%
  \institution{ENSTA ParisTech - U2IS}
  \city{Palaiseau Cedex}
  \country{France}
}

\author{Kostiantyn Potomkin}
\affiliation{%
\institution{Australian National University}
\city{Canberra}
\country{Australia}
}

\author{Christian Schilling}
\affiliation{%
  \institution{IST Austria}
  \city{Klosterneuburg}
  \country{Austria}
}

\renewcommand{\shortauthors}{S. Bogomolov et al.}

\begin{abstract}
We present \textsc{JuliaReach}, a toolbox for set-based reachability analysis of dynamical systems. \textsc{JuliaReach} consists of two main packages: \texttt{Reachability}, containing implementations of reachability algorithms for continuous and hybrid systems, and \texttt{LazySets}, a standalone library that implements state-of-the-art algorithms for calculus with convex sets. The library offers both concrete and lazy set representations, where the latter stands for the ability to delay set computations until they are needed. The choice of the programming language Julia and the accompanying documentation of our toolbox allow researchers to easily translate set-based algorithms from mathematics to software in a platform-independent way, while achieving runtime performance that is comparable to statically compiled languages. Combining lazy operations in high dimensions and explicit computations in low dimensions, \textsc{JuliaReach} can be applied to solve complex, large-scale problems.

\end{abstract}



\maketitle


\section{Introduction}
\label{sec:introduction}
Set-based reachability analysis is a rigorous approach to reason about the behavior of a dynamical system~\cite{AlurCHHHNOSY95}.
Many well-known approaches to reachability analysis are based on flowpipe construction, i.e., computing a finite cover of all trajectories starting from a given set of initial states~\cite{LeGuernicG09,frehse2011spaceex,ChenAS13flowstar,Althoff15cora,bak2017simulation,bogomolov-etal:cav2012,ray2015xspeed}.
Such approaches, while being based on different theoretical ideas, conceptually share underlying machinery.
Thanks to these common grounds, when implementing a flowpipe construction approach (in a prototype or in an end-user tool), our vision is that, instead of starting from scratch, one can reuse the code from a related implementation.
To make such reuse convenient, it is essential to provide a flexible interface for common operations, supported by an efficient library with rich functionality.
Given these features, researchers and their students can then concentrate on the core algorithms and ideally prototype their approach with relatively few lines of code.
In practice, however, many of the existing approaches based on flowpipe construction are implemented in different tools without any shared code base.

An important element in this context is the choice of programming language.
In academic prototyping, the choice is usually a consideration between convenience during development and output performance in the final product.
On the one hand, compiled languages such as \cpp offer high performance, but the compilation overhead is inconvenient for prototyping.
On the other hand, interpreted languages such as \python offer an interactive session for convenient prototyping, but these languages fall behind in performance or need to extend the code to work with another lower-layer program such as \tool{Numba} or \tool{Cython} (known as the two-language problem).
A compromise between the two worlds are just-in-time (JIT) compiled languages such as \matlab.

\julia, a recent%
\footnote{Version 1.0 of \julia was released in August 2018.}
programming language for high-performance scientific computing~\cite{bezanson2017julia},
reconciles the two advantages of compiled and interpreted languages described above, as it comes with an interactive read-evaluate-print loop (REPL) front-end, but is JIT compiled to achieve performance that is competitive with compiled languages such as \cPL.
A distinctive feature of \julia is multiple dispatch (i.e., the function to execute is chosen based on each argument type), which allows to write efficient machine code based on a given type, e.g., of the set.
As additional features, \julia is platform independent, has an efficient interface to \cPL, is supported in \tool{Jupyter} notebooks and well-suited for parallel computing.
\julia has a determined and quickly-growing community, especially for scientific tools.%
\footnote{\url{https://julialang.org}}
We believe that all this makes \julia an interesting programming language for writing a tool for reachability analysis.

\medskip
In this paper, we present \juliareach%
\hide{\footnote{\url{http://juliareach.org}}}, an open-source toolbox for rapid prototyping of set-based reachability approaches, written in \julia.
%
%
The \juliareach toolbox consists of two main packages:
The first package, \reachability, offers core infrastructure for implementing reachability algorithms for continuous and hybrid systems.
As a proof of concept, the package currently provides implementations of an algorithm for the analysis of linear time-invariant (LTI) systems based on decomposition~\cite{HSCC-deco}, and of the classic interleaving algorithm of continuous and discrete-post operators as used in hybrid system tools such as \spaceex~\cite{frehse2011spaceex,bogomolov-etal:hvc2014,bogomolov-et-al:sttt-2015}.

The second package, \lazysets, is a library for state-of-the-art calculus with convex sets.
As the name indicates, a key feature of \lazysets is lazy set representation using support functions.
For making the lazy sets concrete, the library offers means to overapproximate them, e.g., using template directions or $\varepsilon$-close approximation.
\lazysets also supplies concrete set representations, e.g., polyhedra, zonotopes, and ellipsoids.

A binary file to use \julia can be downloaded on the official web page, and installing the \juliareach packages is just one command.
The toolbox puts attention on high efficiency and usability, both from the end-user side and the developer side, i.e., the code is thoroughly tested and documented, including tutorials, and written with extensibility in mind.
Code contributions follow a continuous-integration scheme and are peer-reviewed to assure high quality.

\medskip
\paragraph{Related Work}
Several reachability tools are implemented in \cpp.
For example, \spaceex is a \cpp package that currently integrates several independent reachability algorithms for linear hybrid systems~\cite{frehse2011spaceex}.
\flowstar implements Taylor model approximation for nonlinear hybrid systems in \cpp~\cite{ChenAS13flowstar}.

There are also other reachability tools that are written in JIT-compiled or interpreted languages.
For example, \cora implements several algorithms for linear and nonlinear hybrid systems in \matlab~\cite{Althoff15cora}, and also offers a standalone set library focused on zonotopes, as well as a library for interval arithmetic.
A subset of the algorithms from \cora have been implemented in \cpp in the tools \tool{CORA/SX} and \tool{SymReach}, leading in some instances to a speedup of up to a factor of 3~\cite{ARCH18:nonlinear}.
A few tools are implemented in \python. For instance, \hylaa implements discrete-time reachability algorithms for linear hybrid systems ~\cite{bak2017simulation}.

\hypro is a \cpp library for convex set representation with similar aims as the \lazysets library~\cite{SchuppAMK17hypro}.
It provides a common interface for set representations like boxes, convex polyhedra, support functions, and zonotopes.
Recently, a reachability tool called \tool{HyDRA}~\cite{ARCH18:linear} has been implemented on top of \hypro.
At the time of writing, \tool{HyDRA} is not publicly available.

\medskip

In the remainder of the paper,
we describe the \lazysets library in Section~\ref{sec:lazysets} and the \reachability package in Section~\ref{sec:reachability}.
In Section~\ref{sec:case_study}, we demonstrate the viability of \juliareach in two case studies.

\section{The LazySets Library}
\label{sec:lazysets}
This section outlines the \lazysets library.
In the library we consider representations of closed convex sets in the usual sense from convex geometry:
A set is closed if it contains all its limit points.
A set $S$ is convex if for any $m$ points $v_j \in S$ and $m$ non-negative numbers $\lambda_j$ that sum up to 1 we have that $\sum_j \lambda_j v_j \in S$ as well.
Alternatively, a closed convex set is an intersection of (possibly infinitely many) closed half-spaces.

Every convex set type in the library inherits from the parametric abstract type \code{LazySet\{N\}}, where \code{N} is a parameter for the numeric type.
This way one can easily switch between, e.g., floating point (\code{Float64}) and exact (\code{Rational}) precision with no additional performance penalty:
At runtime, \tool{Julia} uses multiple dispatch on \code{N} and JIT-compiles into type-specific code.

The library comes with a collection of common set representations, such as balls in different norms, ellipsoids, hyperrectangles, polyhedra, polytopes (i.e., bounded polyhedra) in constraint and in vertex representation, and zonotopes.
One key feature of \lazysets is to apply common binary set operations in a lazy fashion, i.e., to not construct the result of an operation explicitly.
For that purpose, we use special wrapper types to represent operations such as convex hull, Minkowski sum, linear and exponential map, intersection, and Cartesian product.
Since these lazy set operations themselves are subtypes of \code{LazySet}, they can also be nested.

New subtypes of \code{LazySet} can be easily added.
To realize the lazy paradigm, a set type should provide an implementation of two functions:
The first function has signature \code{$\sigma$($d$,\,$S$)} in \lazysets and returns some furthest point of a set $S$ in a given direction $d$, i.e., its support vector.
We recall that the set of support vectors corresponds to the optimal points for the support function $\rho$, i.e., $\sigma_{S}(d) := \{x \in S \mid d^\mathrm{T}x = \rho_S(d)\}$, $\rho_S(d) := \max_{x \in S} d^\mathrm{T} x$.
%

The second function $\rho_S(d)$ has signature \code{$\rho$($d$,\,$S$)} in \lazysets and is exactly the support function.
Since the default implementation of the support function is given by the above formula, it suffices to define the function $\sigma$; but if a more efficient implementation is available, it will be used automatically by the dispatch machinery in \julia.
For the above lazy operations, the support vector can be evaluated efficiently without explicitly representing the set resulting from the operation~\cite{LeGuernic2010250,LeGuernic09}, with the exception of intersection.
For intersection, we only provide (over)approximate values for $\rho$, where we can either use a line search algorithm inspired by~\cite{Frehse012} or a coarser heuristics given as $\rho_{\X \cap \Y}(d) \leq \min(\rho_{\X}(d), \rho_{\Y}(d))$~\cite{LeGuernic09}.

\lazysets defines several \emph{abstract supertypes} (sometimes called \emph{interfaces} in other programming languages) to bundle common functionality.
For instance, all polytopic set types should subtype the \code{AbstractPolytope} type and implement methods to obtain their vertex representation and their constraint representation.
This allows to write generic (or ``virtual'') functions for arbitrary sets that belong to the \code{AbstractPolytope} family.
If later a new set type is added to this family, all such generic functions are available.


\subsection{The lazy paradigm}

For illustration, consider the linear map of a set $S$ by a matrix $M$. In \juliareach, we can either write \code{LinearMap($M$,\,$S$)} or use the short hand \code{$M$\,*\,$S$} for convenience.
The \code{LinearMap} type has two fields that hold the map and the set, respectively.
If $S$ is itself a \code{LinearMap} instance, the constructor multiplies the two matrices immediately.
Otherwise, it creates a \code{LinearMap} instance that wraps $M$ and~$S$ instead of computing the map explicitly.
Given a direction $d$, the support vector computation is based on the formula $M \cdot \sigma(M^T d, S)$, i.e., it asks for the support vector of the wrapped set $S$.

As a second example, consider two polytopes $P_1$ and $P_2$. The command \code{$P_1 \oplus P_2$} instantiates a new \code{MinkowskiSum} instance.
Similar to the linear map, the support vector can be defined recursively as $\sigma_{P_1\oplus P_2}(d) = \sigma_{P_1}(d) \oplus \sigma_{P_2}(d)$.
Again, the binary operation \code{MinkowskiSum} is defined between any two \code{LazySet} types.
On the other hand, this operation can be performed explicitly using the command \code{minkowski\_sum($P_1$,\,$P_2$)}, which calls an external library called \code{Polyhedra.jl}~\cite{LegatPolyhedra},%
that provides a unified interface to well-known implementations of polyhedral computations, such as \tool{CDD} or \tool{LRS}.

The approach lazy-by-default and optionally-explicit is also available with other set types.
For example, we can execute the concrete \code{linear\_map} function for \code{Zonotope}{}s, in which case the generators will be transformed.

\smallskip


\medskip

In practice it is convenient to switch from a lazy to a concrete representation at some point, ideally without an exponential increase in the computational cost when dealing with high-dimensional problems.
Below we describe the \approximations sub-module, which serves this purpose.

\subsection{From lazy to concrete set representation}

The \approximations module implements the transfer from a lazy set representation to a concrete set, generally involving an overapproximation.
As a particular feature, the module can be used to combine lazy high-dimensional sets with explicit low-dimensional approximations of projections.
For illustration, consider the following example.
A typical set equation for a discrete approximation model of LTI systems is
\begin{equation}\label{eq:lti_discretization}
    Y = \text{CH}(e^{A\delta} X_0 \oplus \delta B U, X_0),
\end{equation}
with real matrix $A \in \R^{n \times n}$, time step $\delta$, initial states $X_0$, nondeterministic inputs $B U$, and CH denoting the convex hull.
For example, let $\delta = 0.1$, $X_0 \subseteq \R^{n}$ be a ball with center $(1,\dots,1)$ and radius $0.1$ in the infinity norm, $U \subseteq \R^{m}$ be a ball centered in the origin with radius $1.2$ in the 2-norm, and $B$ be a linear map of appropriate dimensions.
With \lazysets, Eq.~\eqref{eq:lti_discretization} for $n=1{,}000$, $m=2$, and random matrix coefficients is

\medskip\noindent\begin{minipage}{\linewidth}
\begin{lstlisting}
 n = 1000; m = 2; $\delta$ = 0.1
 A = sprandn(n, n, 0.01); B = randn(n, m)
 X0 = BallInf(ones(n), 0.1)
 U = Ball2(zeros(m), 1.2)
 Y = ConvexHull(SparseMatrixExp(A*$\delta$)*X0$\,\oplus\,\delta$*B*U,$\,$X0)
\end{lstlisting}
\end{minipage}

The execution is instantaneous because we just created a nested \emph{lazy} set.
Note the use of \code{SparseMatrixExp($M$)}, a wrapper around the matrix exponential $e^M$; the evaluation of the support vector for the lazy matrix exponential relies on computing the \textit{matrix action}, a technique taken from the numerical ODE/PDE domain.

\smallskip


\begin{figure}[t]
	\includegraphics[width=\linewidth,height=7cm,keepaspectratio]{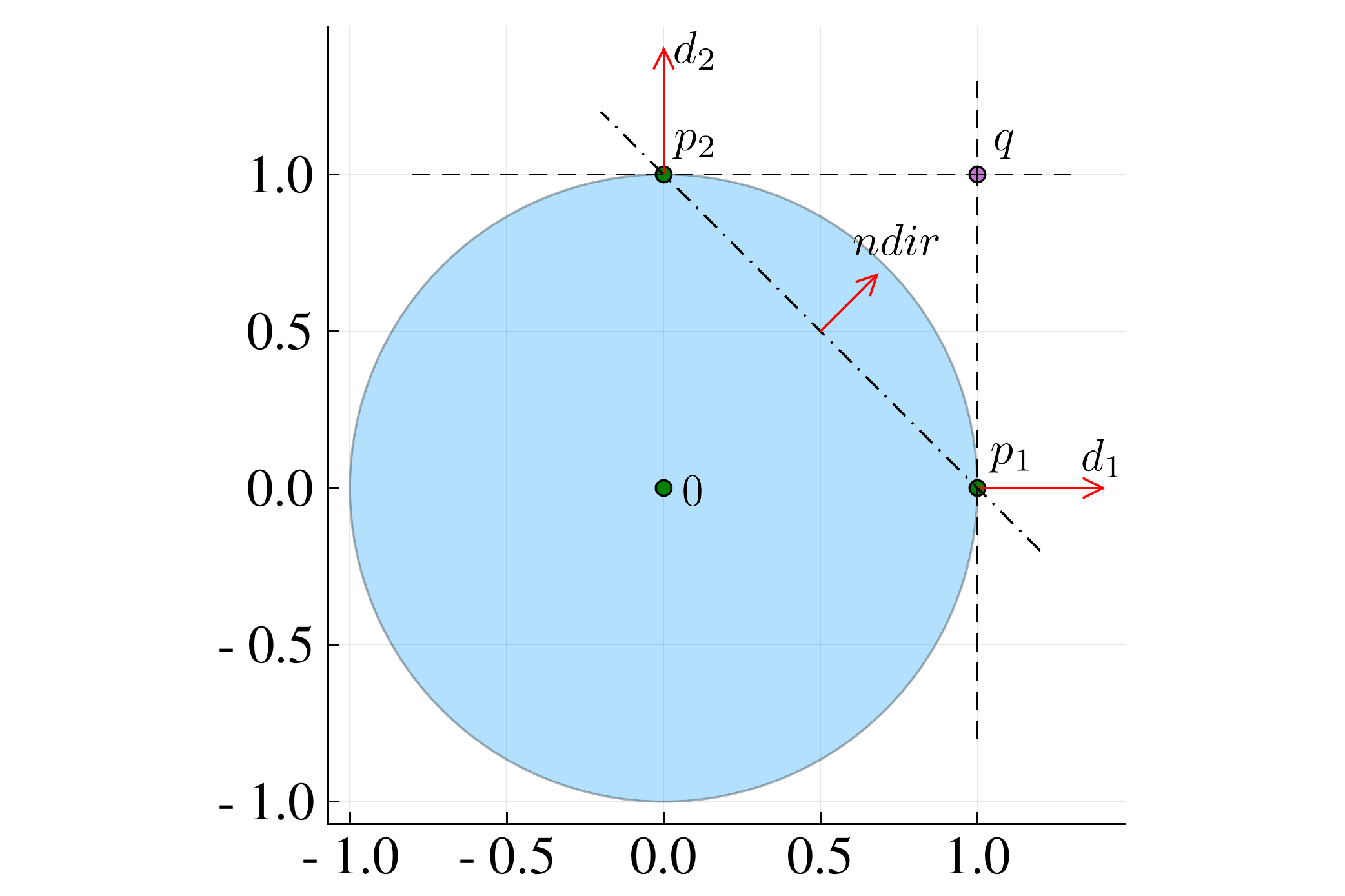}
	\caption{
	Description of the overapproximation algorithm.
	We start with box directions ($\pm d_1$ and $\pm d_2$) and then check for each angle if the distance between the support vector of the original set and the vertex is less than an error bound~$\varepsilon$.
	}
	\label{fig:iterative_refinement}
\end{figure}

The \approximations module can now be used to get information about \code{Y} without having computed any concrete representation; e.g., projection of a set into two dimensions is useful for visualization or verification of properties that involve only two variables.
For this purpose, the module offers the function \code{overapproximate($S$, $\varepsilon$)} where $S$ is a set and $\varepsilon$ is an error bound.
The function overapproximates a 2D set by adding supporting directions until the error bound is achieved (measured in terms of the Hausdorff distance), where the number of directions is optimal.
The algorithm implements Kamenev's method~\cite{Kamenev96,lotov2008modified}, which we sketch in \fig{fig:iterative_refinement}.

\begin{figure}[tb]
	\includegraphics[width=\linewidth,height=4.5cm,keepaspectratio]{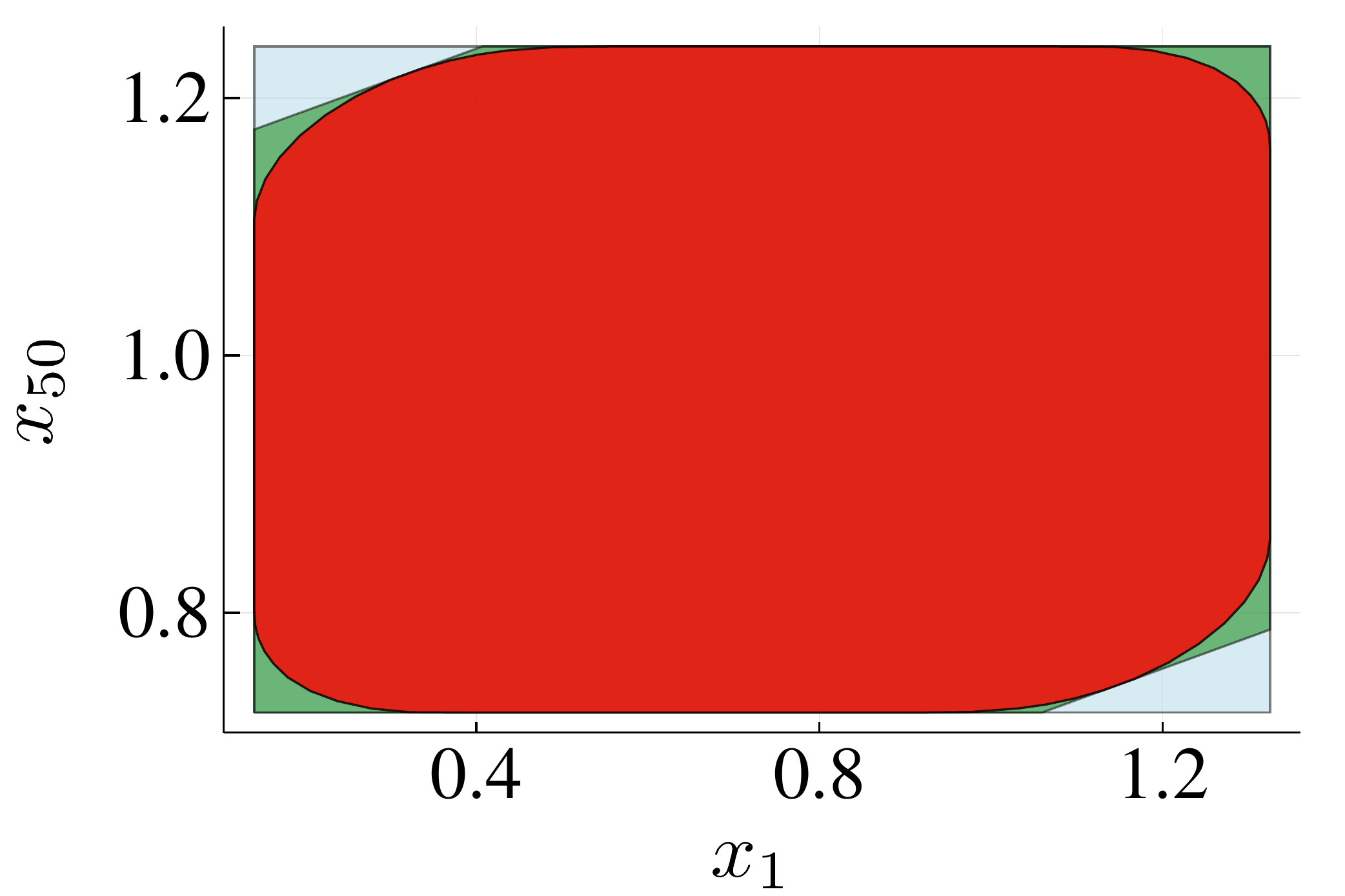}
	\caption{
	Decomposition of the $1{,}000$-dimensional set \code{Y}, projected onto $x_1$ and $x_{50}$, with polytope approximations of different precision: $\varepsilon = \textrm{Inf}$ (box directions, blue), $\varepsilon = 0.1$ (green), and $\varepsilon = 0.001$ (red).
	Runtimes are 55 ms, 195 ms and 2 s, respectively.
	}
	\label{fig:overapproximation}
\end{figure}

Suppose now that we are interested in observing the projection of \code{Y} onto the variables~1 and~50.
First we define the $2 \times n$ projection matrix and apply it as a linear map (i.e., from the left).
Then we use the \code{overapproximate} method with a specified value of $\varepsilon$:

\medskip\noindent\begin{minipage}{\linewidth}
\begin{lstlisting}
 $\pi$ = spzeros(2, n); $\pi$[1, 1] = $\pi$[2, 50] = 1
 res = overapproximate($\pi$*Y, $\varepsilon$)
\end{lstlisting}
\end{minipage}

\fig{fig:overapproximation} shows some results for different values of $\varepsilon$.

For high-dimensional overapproximation, instead of $\varepsilon$ one can alternatively pass a set type (e.g., \code{Hyperrectangle} for box approximation) or a special iterator type for template directions.
We note again that one can call the \code{overapproximate} function with an arbitrary (nested) lazy set, as long as it implements \code{$\rho$($d$,\,$S$)} to evaluate the support function for a given direction.

\section{The Reachability Package}
\label{sec:reachability}
The purpose of the \reachability package is to provide the basic infrastructure for implementing a custom reachability algorithm for continuous and hybrid systems, modeled as a hybrid automaton~\cite{AlurCHH92}.
We assume that the reader is familiar with the basic terminology of hybrid automata.

\subsection{Infrastructure}

\reachability provides the basic infrastructure that is needed or at least convenient to have for writing reachability algorithms.

An \code{Options} type stores user-defined options as key-value pairs that can be passed around to the solver backends.

Dynamical systems are passed to \reachability as special wrapper types.
The reachability algorithms themselves are interfaced with post operators, which we will describe later.
For hybrid systems, it is common to use clustering algorithms and fixed-point checks.
\reachability offers standard convex-hull clustering and checks for a fixed-point (i.e., inclusion) after every jump, either before or after clustering (controlled by the user).

We support post-processing the resulting sets with an output function or a projection to given output dimensions.
The sets can be visualized with the external \tool{Plots.jl} tool using a single command.

\subsection{Continuous-Post Operators}
The interface for continuous-post operators is rather simple.
Based on an \code{:algorithm} option, we just call the respective interface function with the continuous model, a \code{ReachSet} object, and a list of options (the number of steps, the step size, and other algorithm-specific settings).
A \code{ReachSet} is a wrapper of a set and a corresponding uncertain time interval.
Initially we would call the algorithm with the initial states and the time point 0 as in \code{(X0, [0, 0])}, but in the context of a hybrid system, the uncertain time interval will grow with each jump.
The expected return type is just a sequence of \code{ReachSet}{}s, sometimes called a reach tube.

For out-of-the-box usage, we provide an efficient implementation of a recent approach for the reachability analysis of LTI systems~\cite{HSCC-deco,ARCH18:linear}.
The idea is to first decompose the system into low-dimensional \emph{blocks} and then solve many small reachability problems.
While the decomposition comes with an approximation error, the approach is suited for large-scale sparse systems.

\subsection{Discrete-Post Operators}
\label{sec:discrete_post_ops}

For analyzing hybrid systems, in addition to a post operator for continuous systems, one needs a post operator for the discrete transitions.
Given a reach tube \X, a transition with a guard \Gu and assignment $\asgn(\cdot)$, and a target invariant \I, the discrete-post operator should compute (an overapproximation of) the set $\asgn(\X \cap \Gu) \cap \I$ and pass this set on to the continuous-post operator again.

Having both a continuous and a discrete-post operator, one can implement a classic algorithm that essentially interleaves the two post operators.
\reachability offers the function ``\code{solve}'', which implements this algorithm for given post operators.

We implemented two discrete-post operators to instantiate the above algorithm.
The first operator uses (concrete, i.e., non-lazy) polytopes in constraint representation and relies on the external library \tool{Polyhedra.jl} to perform the above-mentioned operations.
In particular, the intersections are computed explicitly.

The second discrete-post operator is lazy but offers the options to overapproximate any of the intersections for obtaining a concrete set.
Since evaluating a nested lazy intersection is a computationally complex task, this operator uses the line-search algorithm (see Section~\ref{sec:lazysets}) only on the highest nesting level, and uses the coarser heuristics on lower nesting levels.
If one is only interested in the support vector in one specific direction, this purely lazy approach scales very well.
For more directions, however, there is no clear recipe which of the intersections should be performed lazily and which one should be overapproximated.

In the next section, we compare the performance of these two operators in a case study.
We will instantiate the lazy discrete-post operator once with both intersections kept lazy and once with both intersections overapproximated.

\section{Case Study}
\label{sec:case_study}
\begin{figure}[tb]
	\centering
	\includegraphics[width=\linewidth,height=6cm,keepaspectratio]{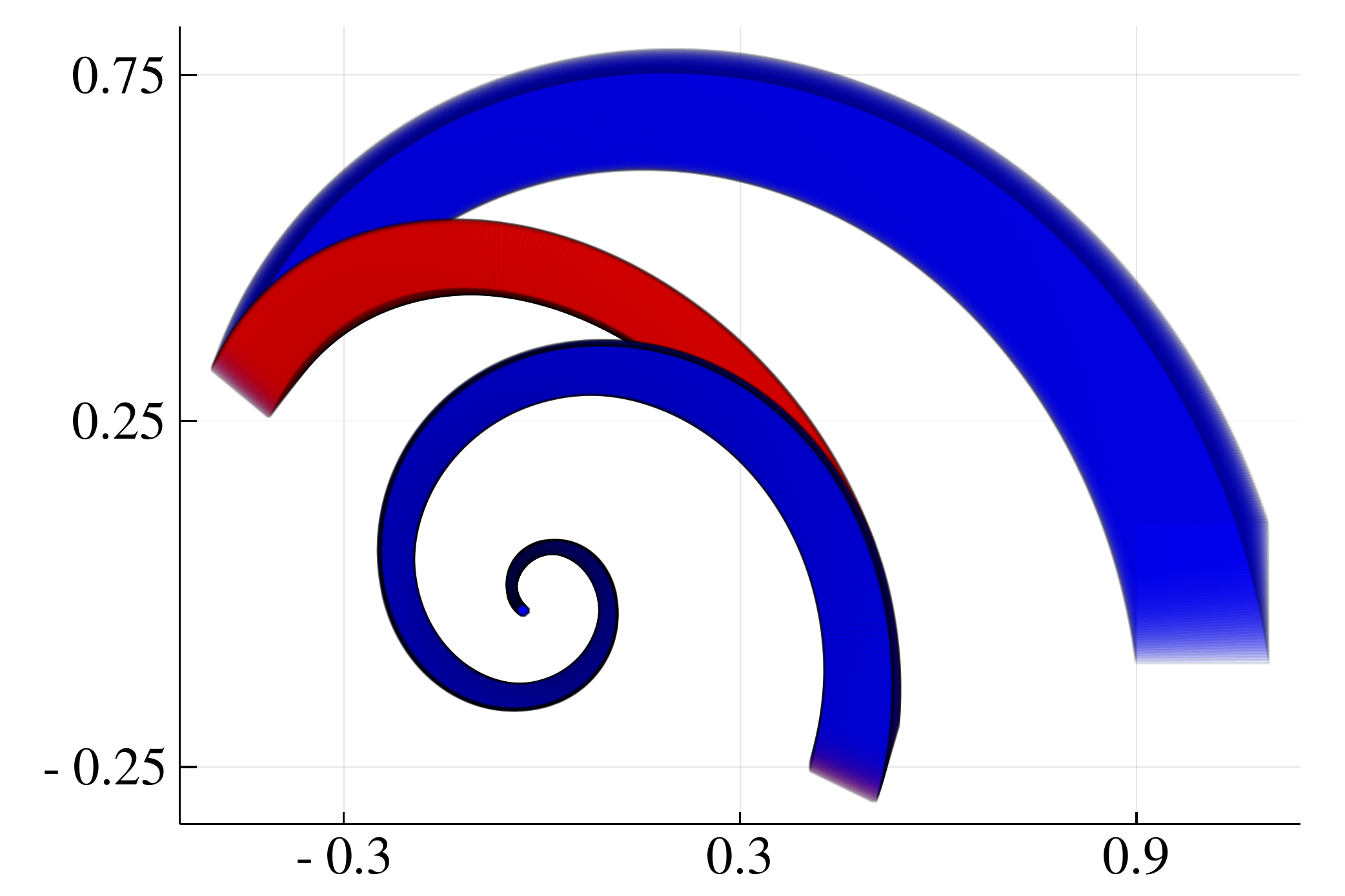}
	\caption{Reachability plot of a two-mode hybrid system. The colors represent location~1 (blue) and location~2 (red), respectively.}
	\label{fig:zonotope_reach}
\end{figure}

We perform two case studies that investigate some applications.
In the first one, we re-implement a classic approach for reachability analysis of LTI systems using the \lazysets library.
We then add a custom extension for hybrid systems, demonstrating the rapid-prototyping capabilities of \julia.
In the second case study, we consider \reachability and compare the performance of different discrete-post operators outlined in the previous section.

\subsection{Working with the LazySets Library}
\label{sec:cs_lazysets}

We wrote a full-fledged re-implementation of Girard's reachability algorithm for LTI systems with zonotopes~\cite{Girard05} in \julia.
The goal is to demonstrate prototyping with \lazysets and the general applicability as a standalone library.
The complete \julia source code is given in Algorithm~\ref{algo:reach_zonotope_continuous}, where we underlined those parts that involve the \lazysets library.

We then extended the example presented in~\cite[Section~4]{Girard05} to a hybrid system of two modes $\ell_i$, $i = 1,2$, with initial states $[0.9, 1.1] \times [-0.1, 0.1]$ and uncertain inputs from a set $u$ with $\mu = \Vert u \Vert_\infty = 0.001$.
The dynamics matrices $A_i$ are defined as follows:
\[
	A_1 = \begin{pmatrix} -1 & -4 \\ 4 & -1 \end{pmatrix} \qquad A_2 = \begin{pmatrix} 1 & 4 \\ -4 & -1 \end{pmatrix}
\]
We add a transition $t_i$ from mode $\ell_i$ to $\ell_{3-i}$ with a hyperplane guard $g_i$: $ g_1 \triangleq x_1 = -0.5$, $g_2 \triangleq x_2 = -0.3$.
\lazysets offers order reduction for zonotopes, which we used here with an upper bound of 10 generators.
We implemented a custom reachability algorithm for this case study, again only using the \lazysets capabilities.
For simplicity, a transition is taken as soon as an intersection with the guard is detected.
\ifappendix{%
The complete source code is given in Appendix~\ref{sec:code_hybrid}.%
}\fi
We plot the reachable states for the time interval $[0, 4]$ and time step $\delta = 0.001$ in \fig{fig:zonotope_reach}.
The analysis takes 0.25 seconds.

\begin{algorithm}[t]
	\caption{Simple LTI reachability algorithm using zonotopes in \julia.}
	\label{algo:reach_zonotope_continuous}
	\DontPrintSemicolon
	\SetKwProg{Fn}{function}{}{end}
	\SetKwFor{For}{for}{}{end}
	\SetKwIF{If}{ElseIf}{Else}{if}{}{elseif}{else}{end}
	\SetKw{In}{in}
	\SetKwData{Xzero}{\lskw{X0}}
	\SetKwData{LazySet}{\lskw{LazySet}\hspace*{-1mm}}
	\SetKwData{Zonotope}{\lskw{Zonotope}\hspace*{-1mm}}
	\SetKwData{ms}{\lskw{minkowski\_sum}\hspace*{-1mm}}
	\SetKwData{lm}{\lskw{linear\_map}\hspace*{-1mm}}
	\SetKwData{ReduceOrder}{\lskw{reduce\_order}\hspace*{-1mm}}
	\SetKwData{and}{\&\&}
	\SetKwData{Times}{\lskw{$*$}}
	\SetKwComment{tcp}{\# }{}
	\Fn{ReachContinuous(A, \Xzero, $\delta$, $\mu$, T, max\_order)}{
		\tcp{bloating factors}
		Anorm = norm(A, Inf)\;
		$\alpha$ = (exp($\delta*$Anorm) $-$ 1 $-$ $\delta*$Anorm)/norm(\Xzero, Inf)\;
		$\beta$ = (exp($\delta*$Anorm) $-$ 1)$*\mu$/Anorm\;
		\BlankLine
		\tcp{discretized system}
		n = size(A, 1)\;
		$\phi$ = exp($\delta*$A)\;
		N = floor(Int, T/$\delta$)\;
		\BlankLine
		R = Vector\{\LazySet\}(N)\;
		\If{N == 0}{
			\Return R\;
		}
		\BlankLine
		\tcp{initial reach set in time interval [0, $\delta$]}
		$\phi$p = (I+$\phi$)/2\;
		$\phi$m = (I$-\phi$)/2\;
		gens = hcat($\phi$p $*$ \Xzero.generators, $\phi$m $*$ \Xzero.center, $\phi$m $*$ \Xzero.generators)\;
		R[1] = \ms(\Zonotope($\phi$p $*$ \Xzero.center, gens), \Zonotope(zeros(n), ($\alpha + \beta$)$*$eye(n)))\;
		\If{order(R[1]) $>$ max\_order}{
			R[1] = \ReduceOrder(R[1], max\_order)
		}
		\BlankLine
		\tcp{recurrence for [$\delta$, 2$\delta$], $\dots$, [(N$-$1)$\delta$, N$\delta$]}
		ball$\beta$ = \Zonotope(zeros(n), $\beta*$eye(n))\;
		\For{i \In 2:N}{
			R[i] = \ms(\lm($\phi$, R[i$-$1]), ball$\beta$)\;
			\If{order(R[1]) $>$ max\_order}{
				R[i] = \ReduceOrder(R[i], max\_order)
			}
		}
		\Return R\;
	}
\end{algorithm}

\subsection{Working with the Reachability Framework}
\label{sec:cs_reachability}

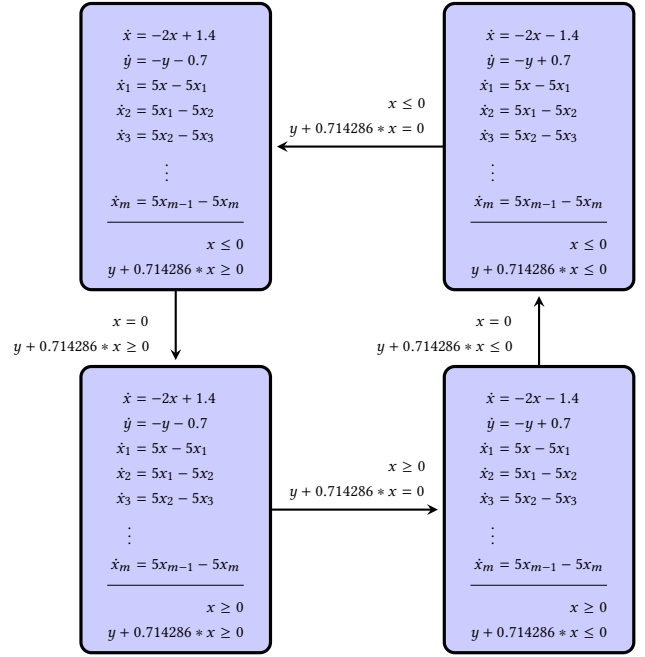
\begin{figure}[t]
	\begin{minipage}{\linewidth}
		\hspace*{-3mm}
		\begin{tikzpicture}[shorten >=2pt,node distance=7cm,on grid,auto,transform shape,scale=.69]
     \node[draw=black, very thick, rectangle, rounded corners, inner sep=5, inner ysep=5, align=center, text width=3.3cm,anchor=base,fill=blue!20] (q_1)   {
     \begin{minipage}{0.8\textwidth}
      \begin{align*}
        \dot{x} &= -2x + 1.4 \\
        \dot{y} &= -y - 0.7 \\
        \dot{x}_1 &= 5x - 5x_1 \\
        \dot{x}_2 &= 5x_1 - 5x_2 \\
        \dot{x}_3 &= 5x_2 - 5x_3 \\
        & \qquad \vdots \\
        \dot{x}_m &= 5x_{m-1} - 5x_m
      \end{align*}
      \begin{align*}
        \hline \\[-2mm]
         x &\leq 0 \\
         y + 0.714286*x &\geq 0
      \end{align*}
       \end{minipage}
     };
     \node[draw=black, very thick, rectangle, rounded corners, inner sep=5, inner ysep=5, align=center, text width=3.3cm,anchor=base, fill=blue!20] (q_2) [right of = q_1] {
     \begin{minipage}{0.8\textwidth}
      \begin{align*}
        \dot{x} &= -2x - 1.4 \\
        \dot{y} &= -y + 0.7 \\
        \dot{x}_1 &= 5x - 5x_1 \\
        \dot{x}_2 &= 5x_1 - 5x_2 \\
        \dot{x}_3 &= 5x_2 - 5x_3 \\
        \vdots \\
        \dot{x}_m &= 5x_{m-1} - 5x_m
      \end{align*}
      \begin{align*}
        \hline \\[-2mm]
         x \leq 0 \\
         y + 0.714286*x \leq 0
      \end{align*}
       \end{minipage}
     };
     \node[draw=black, very thick, rectangle, rounded corners, inner sep=5, inner ysep=5, align=center, text width=3.3cm,anchor=base, fill=blue!20] (q_3) [below of = q_1, node distance=7cm] {
     \begin{minipage}{0.8\textwidth}
      \begin{align*}
        \dot{x} &= -2x + 1.4 \\
        \dot{y} &= -y - 0.7 \\
        \dot{x}_1 &= 5x - 5x_1 \\
        \dot{x}_2 &= 5x_1 - 5x_2 \\
        \dot{x}_3 &= 5x_2 - 5x_3 \\
        \vdots \\
        \dot{x}_m &= 5x_{m-1} - 5x_m
      \end{align*}
      \begin{align*}
        \hline \\[-2mm]
         x &\geq 0 \\
         y + 0.714286*x &\geq 0
      \end{align*}
       \end{minipage}
     };
     \node[draw=black, very thick, rectangle, rounded corners, inner sep=5, inner ysep=5, align=center, text width=3.3cm,anchor=base, fill=blue!20](q_4) [right of = q_3] {
     \begin{minipage}{0.8\textwidth}
      \begin{align*}
        \dot{x} &= -2x - 1.4 \\
        \dot{y} &= -y + 0.7 \\
        \dot{x}_1 &= 5x - 5x_1 \\
        \dot{x}_2 &= 5x_1 - 5x_2 \\
        \dot{x}_3 &= 5x_2 - 5x_3 \\
        \vdots \\
        \dot{x}_m &= 5x_{m-1} - 5x_m
      \end{align*}
      \begin{align*}
        \hline \\[-2mm]
         x &\geq 0 \\
         y + 0.714286*x &\leq 0
      \end{align*}
       \end{minipage}
     };
      \path[->,>=stealth,thick]
      (q_2) edge node[above] {
      \begin{minipage}{0.4\textwidth}
       \begin{align*}
         x &\leq 0 \\
         y + 0.714286*x &= 0
         \end{align*}
      \end{minipage}
           } (q_1)
      (q_4) edge node[left] {
      \begin{minipage}{0.4\textwidth}
       \begin{align*}
         x &= 0 \\
         y + 0.714286*x &\leq 0
         \end{align*}
      \end{minipage}
      } (q_2)
      (q_3) edge node
      {
      \begin{minipage}{0.4\textwidth}
       \begin{align*}
         x &\geq 0 \\
         y + 0.714286*x &= 0
         \end{align*}
      \end{minipage}
      } (q_4)
      (q_1) edge node[left]
      {
      \begin{minipage}{0.4\textwidth}
       \begin{align*}
        x &= 0 \\
        y + 0.714286*x &\geq 0
         \end{align*}
      \end{minipage}
      }  (q_3);
\end{tikzpicture}
	\end{minipage}
    \caption{Hybrid automaton model of a 2-dimensional oscillator and an $m$-dimensional filter.}
    \label{fig:ha_fo}
\end{figure}

In the second case study, we show how the \reachability framework can be used to quickly evaluate a new approach to reachability analysis.
Consider the case where a researcher came up with a new strategy for the discrete-post operator and now wants to compare to existing approaches.
The researcher would just need to implement the different approaches as different post operators in the \reachability package and then evaluate the performance.
Here we simulate this situation by using the operators that currently exist in the \reachability package (see Section~\ref{sec:discrete_post_ops}).

We compare the performance of the different discrete-post operators using a model of a filtered oscillator~\cite{frehse2011spaceex}.
The model represents a parameterized hybrid system consisting of ($i$) a two-dimensional switched oscillator in the variables $x$ and $y$, and ($ii$) a filter with $m$ state variables $x_1, \dots, x_m$.
We show the hybrid automaton in \fig{fig:ha_fo}.
The filter smooths $x$ with $x_m$ as an output signal, and the amplitude decreases with increasing dimension of the filter.
The initial set is given as $x \in [0.2, 0.3], y \in [-0.1, 0.1]$, and $x_j = 0$ for all $x_j$.

\begin{table*}
    \caption{Runtimes (in seconds) for different discrete-post operators in the \reachability package and the \spaceex tool.}
    \label{tab:d_post_table}
    \begin{center}
      \begin{tabular}{l P{2} P{2} P{2} P{2} P{2} P{2} P{2} P{2} P{2} }
        \hline
        Number of filters $m$ & \hd{2} & \hd{4} & \hd{8} & \hd{16} & \hd{32} & \hd{64} & \hd{128} & \hd{196} & \hd{256} \\
        \hline
        \hline
        \reachability\ -- concrete & 0.60 & 0.84 & 1.77 & 4.24 & 12.18 & 39.23 & 137.89 & 341.38 & 633.86 \\
        \hline
        \reachability\ -- lazy & 0.55 & 0.75 & 1.41 & 3.19 & 6.39 & 22.65 & 64.63 & 145.49 & 262.91 \\
        \hline
        \reachability\ -- lazy with overapproximation & 0.57 & 0.75 & 1.17 & 4.76 & 6.18 & 20.05 & 47.78 & 103.12 & 212.10 \\
        \hline
        \hline
        \spaceex & 0.11 & 0.22 & 0.11 & 0.30 & 1.10 & 6.78 & 56.72 & 262.88 & 1{,}011.17 \\
        \hline
    \end{tabular}
    \end{center}
\end{table*}

As a reference tool, we use \spaceex (in version 0.9.8f) to evaluate the overall performance.
To compare only the performance of the discrete-post operators, we fix the number of jumps to four, which corresponds to one loop in the automaton.
\reachability offers a simple option for that purpose, but \spaceex does not.
To have a fair comparison, we do not use our option and instead modify the model in the following way.
We add a new variable $b$ (for \emph{bound}) with initial value 1.
On the first transition (at the bottom in \fig{fig:ha_fo}) we add an assignment $b := 2*b$.
Finally, we augment the invariant of the target location (lower right) by the constraint $b \leq 2$.

\begin{figure}[tb]
    \includegraphics[width=.495\linewidth,keepaspectratio]{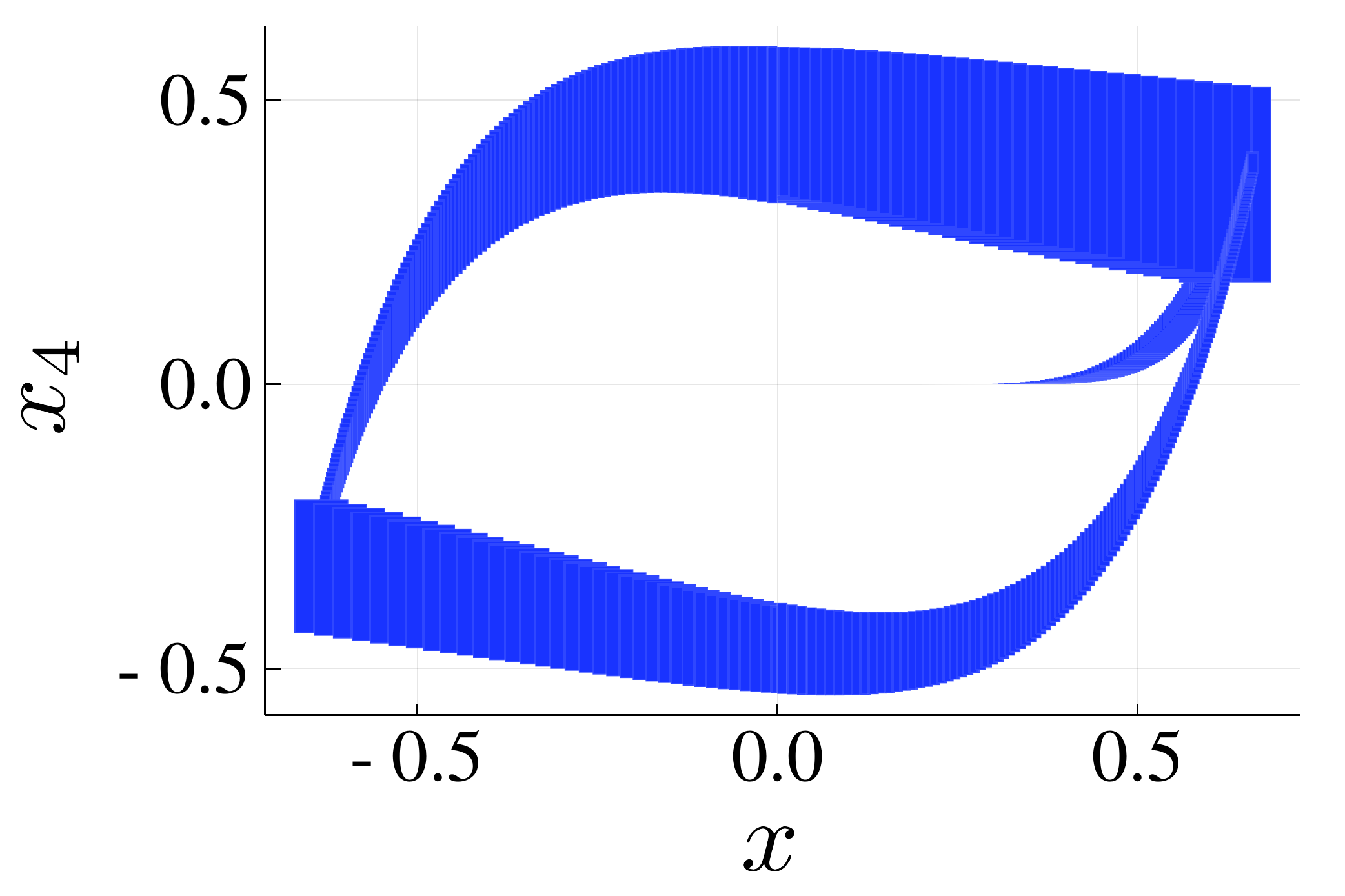}
    \hfill
    \includegraphics[width=.495\linewidth,keepaspectratio]{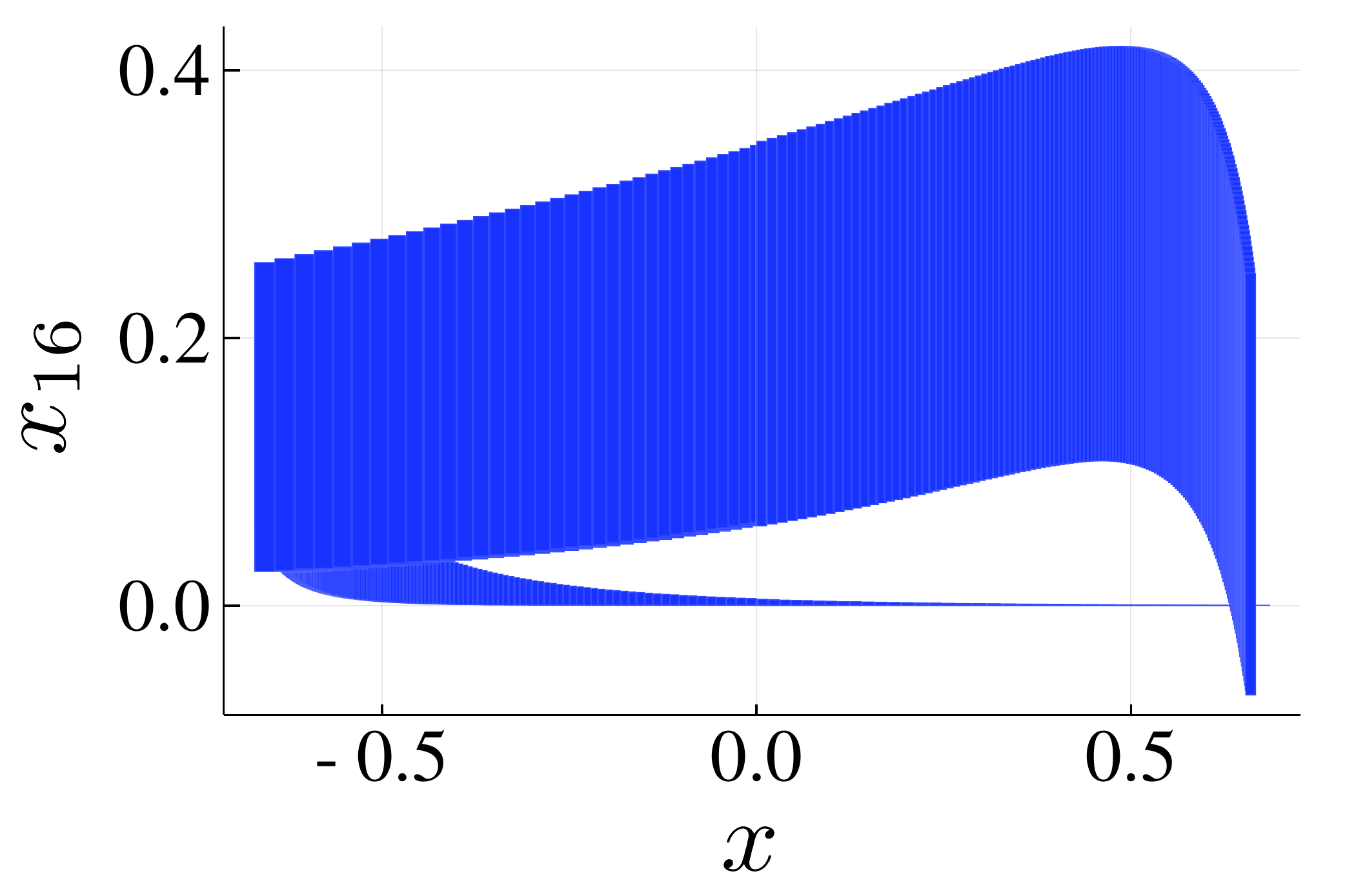}
    \caption{Output for the filtered oscillator model with different filter dimension $m$, using \reachability with the lazy discrete-post operator with overapproximation. Left: $m=4$. Right: $m=16$.}
    \label{Fig:pic/fo4dim}
    \label{fig:fo_hybrid_reach}
\end{figure}

We assume the following benchmark settings.
The time interval is $[0, 20]$ for filters of dimension up to four, and $[0, 99]$ for higher dimensions.
The time step is $\delta = 0.01$ for both \reachability and \spaceex.
We ran the benchmarks on a computer with a 2.20~GHz CPU and 8~GB RAM.
Table~\ref{tab:d_post_table} outlines the performance of the different discrete-post operators and \spaceex.
For \spaceex, we used the support-function algorithm (``\code{supp}'').
\fig{fig:fo_hybrid_reach} shows some plots of the output variables for the lazy discrete-post operator with overapproximation.
(We used box approximations in this evaluation.)

As expected, the concrete intersection with polytopes is the slowest.
(For this model, all computations could be performed in constraint representation.)
The lazy-intersection approaches (second and third row in the table) scale much better, both roughly in the same order.
Thanks to the fast continous-post algorithm, all operators outperform \spaceex for the biggest model instances.

\section{Conclusion}
\label{sec:conclusion}
We have presented the \juliareach toolbox, a new environment for developing reachability algorithms for dynamical systems.
\juliareach is written in the modern programming language \julia that unifies rapid prototyping and high performance.
The toolbox consists of a framework for reachability analysis (\reachability) that is built around a standalone library for calculus with convex sets (\lazysets).

In the future, we will extend the amount of post operators offered in \reachability.
In particular, we are working on a new discrete-post operator that exploits the low-dimensional output of the continuous-post operator in the hybrid loop.

\begin{acks}
	This work was partially supported by the
	\grantsponsor{1}{Air Force Office of Scientific Research}{http://www.wpafb.af.mil/afrl/afosr/}
	under award number\
	\grantnum{1}{FA2386-17-1-4065},
	by the ARC project
	\grantnum{2}{DP140104219}
	(\grantsponsor{2}{Robust AI Planning for Hybrid Systems}{https://cs.anu.edu.au/research/research-projects/robust-ai-planning-hybrid-systems}),
	by the
	\grantsponsor{3}{Austrian Science Fund (FWF)}{https://www.fwf.ac.at/}
	under grants
	\grantnum{3}{S11402-N23} (RiSE/SHiNE)
	and
	\grantnum{4}{Z211-N23} (Wittgenstein Award),
	and the
	\grantsponsor{5}{European Union's Horizon 2020 research and innovation programme}{https://ec.europa.eu/programmes/horizon2020/}
	under the
	Marie Sk{\l}odowska-Curie grant agreement
	No \grantnum{5}{754411}.
	Any opinions, findings, and conclusions or
	recommendations expressed in this material are those of the authors
	and do not necessarily reflect the views of the United States Air
	Force.
\end{acks}

\bibliographystyle{ACM-Reference-Format}
\bibliography{bibliography}

\ifappendix{%
\newpage
\appendix
\section{Hybrid reachability implementation}
\label{sec:code_hybrid}
The following \julia function is an embedding of Algorithm~\ref{algo:reach_zonotope_continuous} in a simple algorithm for hybrid systems.
Apart from \lazysets, no other library is required.
We underlined the places where \lazysets is used.

\begin{algorithm}[h]
	\caption{Simple hybrid reachability algorithm in \julia.}
	\label{algo:reach_zonotope_hybrid}
	\DontPrintSemicolon
	\SetKwProg{Fn}{function}{}{end}
	\SetKwFor{For}{for}{}{end}
	\SetKwIF{If}{ElseIf}{Else}{if}{}{elseif}{else}{end}
	\SetKw{In}{in}
	\SetKw{Break}{break}
	\SetKwData{LazySetN}{\lskw{LazySet}}
	\SetKwData{IsIntersectionEmpty}{\lskw{isdisjoint}\hspace*{-1mm}}
	\SetKwData{and}{\&\&}
	\SetKwComment{tcp}{\# }{}
	\Fn{ReachHybrid(As, Ts, init, $\delta$, $\mu$, T, max\_order)}{
		\tcp{start with initial states at time 0}
		queue = [(init[1], init[2], 0.)]\;
		\BlankLine
		res = Tuple\{\LazySetN, Int\}[\,]\;
		\While{!isempty(queue)}{
			init, loc, t = pop!(queue)\;
			\BlankLine
			\tcp{compute continuous successors}
			R = ReachContinuous(As[loc], init, $\delta$, $\mu$, T-t, max\_order)\;
			\BlankLine
			found\_transition = false\;
			\For{i \In 1:length(R)-1}{
				S = R[i]\;
				push!(res, (S, loc))\;
				\BlankLine
				\tcp{check intersection with guards}
				\For{(guard, tgt\_loc) \In Ts[loc]}{
					\If{!\IsIntersectionEmpty(S, guard)}{
						\tcp{nonempty intersection with a guard}
						new\_t = t + $\delta$ $*$ i\;
						push!(queue, (S, tgt\_loc, new\_t))\;
						found\_transition = true\;
					}
				}
				\BlankLine
				\tcp{stop for first intersection}
				\If{found\_transition}{
					\Break\;
				}
			}
			\If{!found\_transition \and length(R) $>$ 0}{
				push!(res, (R[end], loc))\;
			}
		}
		\Return res\;
	}
\end{algorithm}

}\fi

\end{document}